\documentclass[preprints,article,accept,moreauthors]{Definitions/mdpi}

\firstpage{1} 
\pubvolume{xx}
\issuenum{y}
\articlenumber{z}
\pubyear{2019}
\copyrightyear{2019}
\history{}

\pdfoutput=1

\usepackage{amssymb,amsmath,amsthm,graphicx,hyperref}

\Title{Atom-Field Interaction: From Vacuum Fluctuations to \\ Quantum Radiation and Quantum Dissipation / Radiation Reaction}

\Author{Jen-Tsung Hsiang$^{1}$\orcidA{}, Bei-Lok Hu$^{2}$\orcidB{}}%
\address{%
$^{1}$ \quad Center for High Energy and High Field Physics, National Central University, Chungli 32001, Taiwan; cosmology@gmail.com\\
$^{2}$ \quad Joint Quantum Institute and Maryland Center for Fundamental Physics,\\ University of Maryland, College Park, Maryland 20742, USA; blhu@umd.edu}

\corres{Correspondence: cosmology@gmail.com}

\abstract{
In this paper we dwell on three issues: 1) revisit the relation between vacuum fluctuations and radiation reaction in atom-field interactions, an old issue that began in the 70s and settled in the 90s with its resolution recorded in monographs; 2) the \textit{fluctuation-dissipation relation} (FDR) of the system, pointing out the differences between the conventional form in linear response theory (LRT) assuming ultra-weak coupling between the system and the bath, and the FDR in an equilibrated final state, relaxed from the nonequilibrium evolution of an open quantum system; 3) \textit{quantum radiation} from an atom interacting with a quantum field: We begin with vacuum fluctuations in the field acting on the internal degrees of freedom (idf) of an atom, adding to its dynamics a stochastic component which engenders quantum radiation whose backreaction causes \textit{quantum dissipation} in the idf of the atom. We show explicitly how different terms representing these processes appear in the equations of motion. Then, using the example of a stationary atom, we show how the absence of radiation in this simple cases  is a result of complex cancellations, at a far away observation point, of the interference between emitted radiation from the atom and the local fluctuations in the free field. In so doing we point out in Issue 1 that the entity which enters into the duality relation with vacuum fluctuations  is not radiation reaction, which can exist as a classical entity, but quantum dissipation. Finally, regarding Issue 2, we point out for systems with many atoms,  the co-existence of a set of correlation-propagation relations describing how the correlations between the atoms are related to the  propagation of their (retarded non-Markovian) mutual influence manifesting in the quantum field.   The CPR is absolutely crucial in keeping the balance of energy flows between the constituents of the system, and between the system and its environment. Without the consideration of this additional relation in tether with the FDR,  dynamical self-consistency cannot be sustained. Combination of these two sets of relations forms a generalized matrix FDR relation which capture the physical essence of the interaction between an atom and a quantum field at arbitrary coupling strength.}

\keyword{Vacuum Fluctuations; Quantum Radiation; Quantum Dissipation; Quantum Fluctuation-Dissipation Relation}

\begin{document}

\baselineskip=18pt
\allowdisplaybreaks

\section{Introduction}

In this paper we address/redress several fundamental issues related to the quantum vacuum~\cite{MilBook} in the context of atom-field interactions~\cite{PassanteBook,CTBook,ScullyBook} pertaining to quantum radiative processes. 

 Vacuum fluctuations of a quantum field, described as quantum noise in the environment,  impart a stochastic component in the dynamics of the system,  here, the internal degrees of freedom (idf) of an atom/detector,  interacting with a quantum field.  {\it Quantum dissipation} ensues~\cite{JH1}, obeying a {\it quantum} fluctuation-dissipation relation (FDR)\footnote{Note two sets of FDRs are involved. When we address the vacuum fluctuations vs radiation reaction, we refer to the FDR of the free environment field (in its initial state). However, when we compare the FDR in linear response theory  vs nonequilibrium dynamics, it is the FDR of the system (in its final equilibrium state).} of the field~\cite{RHA,RHK,CPR2D,CPR4D}, on balance with the quantum noise associated with the vacuum fluctuations in the field.  Under specific conditions to be expounded later,  radiation of a quantum nature is emitted from the atom.  This paper highlights the key issues across this spectrum from vacuum fluctuations in a quantum field to stochastic dynamics of the idf of a moving atom/detector to  emitted quantum radiation from it.  We divide these issues into two groups.  
 
 {\bf Group A} is about the relation of vacuum fluctuation and quantum dissipation, which is often misconstrued as radiation reaction of classical radiation theory.  It is the dissipation in the quantum dynamics of atomic system's internal degree of freedom engendered by the vacuum fluctuations of the environmental field. Vacuum fluctuation is intrinsically quantum in nature while radiation reaction  exists at the classical level.  When referring to this dualism, for conceptual clarity, we suggest {\it replacing  the term `radiation reaction'  by quantum dissipation}, or adding the word  `quantum' to it.  2) Vacuum fluctuations are related to quantum dissipation by a fluctuation-dissipation relation.  This relation does not connect  vacuum fluctuations  with classical radiation reaction. Vacuum fluctuations can be viewed (and for Gaussian fields, equated with, by way of the Feynman-Vernon functional identity) as quantum noise.  What spanned two decades of debates (from the early 70s to 90s~\cite{Knight,Milonni,DDC}) mixed with some degree of confusion on this issue stems from the protagonists not recognizing that a distinct level of structure, namely, the stochastic,  is organically tied to the quantum. Many introduced noise as  an {\it ad hoc} entity added in by hand to the classical narrative.  To avoid confusion it is important to place each issue under discussion in its appropriate level of theoretical structure.  This theme is discussed in Sec. II.   Group A issues were already expounded in the work of Johnson and Hu~\cite{JH1,JHFoP}.  We lay out these issues here for the sake of conceptual clarity,  leaving much details to these earlier work.  The main purpose of this paper is to  develop the FDR in two aspects: a)  Showing the differences between  FDRs defined in the context of the nonequilibrium (NEq) dynamics of open quantum systems and FDRs defined conventionally in  linear response theory (LRT). b) the existence of a FDR for a system {\it strongly coupled} with its environment, here represented by an atom interacting with a quantum field.  This is discussed in Section III.  
 
{\bf Group B}  issues are 3) non-Markovian effects between two detectors, and  4) the nature and origin of quantum radiation.  On 3) between pairs of detectors, we show the necessity of including a companion relation, the correlation-propagation relation (CPR) of the field describing how the correlations between the detectors (atoms) are related to the propagation of their (retarded non-Markovian) mutual influence manifesting in the quantum field.  It augments the FDR into a set of generalized  matrix FDRs which capture completely the interaction between an atom and a quantum field at arbitrary coupling. On 4) We focus on how the quantum dynamics of the idf of the atom is related to emitted quantum radiation.  Two noteworthy points are: it is  {\it quantum} radiation, not classical, and it is {\it emitted} radiation, detectable at the far field zone a distance from the atom,  not thermal radiance from a black body, as sensed by a uniformly accelerated detector (UAD) in the Unruh effect. In fact we shall focus on the simplest possible set up, that of a stationary atom in a quantum field, thus there is no Unruh effect~\cite{Unr76} involved.   

Quantum radiation is  different by nature and origin (from quantum interference)  from classical radiation, which can be viewed as the emitted quanta in the limiting condition of a coherent state. Quantum radiation from an accelerated electron under strong external electric field from  a high power laser is described by Sch\"utzhold et al~\cite{Sch06,Sch08}.  Note the characteristics of quantum radiation from an accelerated electron -- the photons are created in pairs (squeezed state) whose polarizations are perfectly correlated,  engendering a nonthermal spectrum.  They  are very different from the Unruh effect proper, albeit both are quantum in nature. Recently  Landulfo, Fulling  and  Matsas~\cite{Fulling19} found a relation between Unruh  thermal radiance in, and Larmor radiation emitted by, a uniformly accelerated charge, the latter  can be seen as entirely built from zero-Rindler-energy modes,  thus bridging the quantum and classical.  As will be seen below, the conceptual scheme here will be  on the level between the quantum and the semiclassical, namely, stochastic effects of vacuum fluctuations playing the role of quantum noise, and their backreaction on the idf of the atom.  Technically, here everything is done at the quantum field theory level. There is no need to introduce noise or stochasticity.  In this paper we shall restrict our attention to stationary harmonic atoms,  known as Unruh-DeWitt detectors in a relativistic context, not to moving charges.  
 
 One way to decipher whether there is quantum radiation is to calculate the stress energy tensor of the quantum field.  Since this is a physical quantity defined in all space, if it vanishes then one can safely conclude there is no emitted radiation.  This was done for a proof of the absence of emitted radiation from a UAD in a 2D spacetime~\cite{HRCapri},  as first suggested by Grove \cite{Grove} and investigated by several groups of authors, e.g.,~\cite{RSG,Unr92,Hin93,MPB,FOC} (for a brief history of  this theme and references in the first decade of its development, see, e.g., the Introduction of~\cite{RHA}.)   Here, with a stationary atom, no radiation of any kind is expected.  We use this simple fact to illustrate how the different contributing parts of energy sources add up to cancel each other.  It is instructive to dissect this problem into parts so we can see the quantum and stochastic components of  the atom, the field and their interplay.   We shall present the results of a calculation of the power  delivered by  different sources, e.g., emitted energy in the quantum radiation generated by the stochastic component in the  dynamics of the system (the idf of the atom) induced by vacuum fluctuations of the quantum field,  is balanced by the vacuum fluctuations at all spatially points far away from the source,  to give us the anticipated result of net zero energy in the far field region.  This is presented in Section \ref{S:etkjssd}.   Quantum radiation from a uniformly accelerated atom  will be treated in a companion paper~\cite{QRadUAD}.

 \section{Theoretical Constructs}
 
It is convenient to address these three aspects: vacuum fluctuations, radiation reaction/quantum dissipation and quantum radiation using three levels of theories --quantum, stochastic and semiclassical -- as described in~\cite{JH1,HJCapri,JHCapri}.  First we show the existence of a fluctuation-dissipation relation linking vacuum fluctuations with quantum dissipation, not classical radiation reaction.  We mention in passing the fundamental difference between thermal radiance in a uniformly accelerated detector (UAD) -- the celebrated Unruh effect~\cite{Unr76} -- and emitted radiation from a moving charge or atom. We then focus on quantum radiation of a very different nature and origin from classical  (Larmor) radiation -- the differences  show up in photon statistics,   spectral and angular distributions~\cite{Sch06}.  Because quantum radiation is weaker by at least an order of $\hbar$, for its detection~\cite{CheTaj}, all the more one needs to know the particular signatures of its existence in the midst of  classical radiation. We also distinguish between the case of a charge, like an electron, and the case of a neutral atom, which in relativistic context is often called a detector.  We work with the so-called Unruh-DeWitt detector~\cite{Unr76,DeW79} modeled by a harmonic oscillator.
 
 \subsection{Classical, Quantum and Stochastic}
 
 {\bf Classical} (Lamor) radiation back-reacts on the moving atom or charge  giving rise to classical {\it radiation reaction}, leading to alternation of its state or trajectories. This  well-known effect is described in standard textbooks~\cite{Jackson} at the classical level and we urge it to stay at that level.  A popular wisdom purports that  vacuum fluctuations  are responsible for radiation reaction, thus one can choose to interpret it either way.  This is sometimes used by ``dualists'' as an example that  one can always find a classical explanation for any quantum phenomena.  Convenient as it may sound this folklore  generates unnecessary confusion because these two entities belong to two different worlds-- vacuum fluctuations are quantum in nature but  radiation reaction already exists at the classical level.  A simple fact debunks this myth:  A charge, be it moving or stationary,  feels the effect of vacuum fluctuations, as would any physical object in all circumstances,  but radiation reaction is absent for a uniformly accelerated (UA) charge.  This can be demonstrated explicitly by use of the Abraham-Lorentz-Dirac equation (e.g., as argued and shown in~\cite{JHFoP}).  We urge saving the term ``radiation reaction'' for classical processes, as in classical electromagnetism and gravitation theory, and not using it in a  quantum context. For a quantum system  to appear classical the process of decoherence is required, and noise in the system's environment (here the quantum field) is instrumental in this. Therefore at the interface between the quantum and (semi-) classical theoretical levels rests the stochastic level. This is well-recognized. The difference is, in conventional treatments one often adds a stochastic source, e.g., noise, to the classical equations of motion and the noise is regarded as a statistical component attached to the classical construct.  This may explain why the aforementioned (misplaced) duality matches (\textit{quantum}) vacuum fluctuations with  (\textit{classical}) radiation reaction. 
 
 {\bf Quantum.}   Turning to the effects of a purely quantum nature, we wish to bring up three issues: Issue  1) is {\it quantum dissipation},  the reactive effect on the atom due to the backreaction of the vacuum  fluctuations of the quantum field on the atom or charge~\cite{JH1} even when it is at rest. For  a moving atom, issue 2) is the \textit{Unruh effect}~\cite{Unr76} in a uniformly accelerated atom.  3) \textit{Quantum radiation}, a subtle entity which is the main focus of this paper.    
 
 On Issue 1,  the relation between vacuum fluctuations and ``radiation reaction'',  to avoid the confusion in mixing up the effects at the classical and the quantum levels we  reiterate the suggestion by Johnson and Hu~\cite{JH1,HJCapri}  that the term ``radiation reaction'' when referring to effects at the quantum level -- such as when the physical observables are treated as operators -- be called ``quantum radiation reaction" or replaced by  {\it quantum dissipation}.   This quantum dissipative reactive effect, not the classical radiation reaction,  is the part which can be considered on the same footing as vacuum fluctuations, when one regards  them as ``two sides of the same coin''. As noticed earlier (e.g.,~\cite{Mil88,AudMul94}) the underlying reason for this  connection is the existence of a fluctuation-dissipation relation between the quantum fluctuations of the field (the environment) and the quantum reactive dissipation in the internal degree of freedom of the atom (the system), an ingrained relation of backreaction in quantum open system dynamics~\cite{RHA,RHK}. ``Fluctuation'' refers to the vacuum fluctuations represented by the Hadamard function (expectation of the anti-commutator of the field) and ``dissipation'' represented by the retarded Green function (expectation of the commutators of the field) refers to the  reactive effect of the field on the atom's idf dynamics.  
 
 {\bf Stochastic}.  Usually a FDR  manifests at the stochastic level, as pointed out in~\cite{JH1,HJCapri},   because when we represent vacuum fluctuations as quantum noise (e.g., using the Feynman-Vernon functional identity  for Gaussian systems) and uses a stochastic equation of motion (Langevin,  master or Fokker-Planck equation) to describe the open system (atom's idf)  we are operating at the stochastic level between the quantum and the semiclassical.   Semiclassical equations of motion are obtained from averaging over the stochastic distributions.  This separation is not academic, the  difference is physical. Here  we show in a quantum operator language the existence of the FDR. There is no need to `convert' the vacuum fluctuations to  quantum noise. As was shown  in~\cite{RHK} for linear systems, the fluctuations and dissipation are represented by the Hadamard and retarded Green's functions.
  
\section{FDR in NEq Dynamics vs Liner Response Theory}

Since the fluctuation-dissipation relation (FDR) is the centerpiece in both sets of issues we indicated in the Introduction, namely, A) the duality between vacuum fluctuations and quantum dissipation / \textit{quantum} radiation reaction; and B) the origin and nature of quantum radiation, as a sum of two parts (shown explicitly  in the next section): one purely originating from the stochastic dynamics of the source (idf of the atom), and  {the other as the consequence of the interference between the radiated field from the source} and the vacuum fluctuations of the field at the point of detection, it is useful to present a fuller description of the features, scopes and implications of FDRs. 

 Whilst FDR is traditionally presented in the context of linear response theory (LRT)~\cite{CalWel,Kubo66,KuboBook,KadBay,FetterWalecka,Fo17} of systems in stationary configurations under weak disturbance, we feel that a new perspective of FDR versed in the context of nonequilibrium (NEq) dynamics is desirable to meet the new challenges of our time, when real time measurements of experimental results are becoming available which enable a broader scope in our understanding of the system's properties.  This is the  approach we have adopted in treating a range of problems from atom-field interactions to quantum processes in black holes and the early universe.  (For a glimpse of the scope of problems whose essences FDR can help to capture, see~\cite{Sciama} and the Introduction of~\cite{CPR2D,CPR4D}.)

\subsection{Differences in the set-ups and the main features}

\noindent\underline{Differences in the set-ups}

Conventional FDR under LRT operates under the following assumptions: the system of interest i) has been in equilibrium with the bath (say, in the Gibbs state) for a sufficiently long time that a stationary condition is established, and ii) is subjected to a weak disturbance and its responses recorded.  By contrast, in  the nonequilibrium (NEq) formalism, the system can be in any arbitrary state, far from the thermal state at the bath temperature $\beta^{-1}$ or the equilibrium state the system finally settles in. Once the initial state of the system  and the properties of the bath are given, for any specified system-bath interaction, we let go of both and let their interaction   determine  the outcome at late times.  The ensuing dynamical evolution of the system under the  influence of fluctuations from the bath can be captured by different methods in NEq dynamics of open quantum systems. The commonly used approaches include the quantum Langevin equation~\cite{FOCqleq} (for detector-field systems, see, e.g,~\cite{LH06} and references therein) or influence functional methods~\cite{FeyVer} (see~\cite{CalHu08} and references therein). Fluctuations in the quantum field lead to quantum dissipation in the atom's idf.  After the system relaxes to a final  equilibrium state their interlocked behavior is captured by the FDR.\\ 

\noindent\underline{Main features: FDR is an emergent relation in NEq} 

Since in LRT the FDR is formulated in terms of perturbative theory with respect to the equilibrium state of the system, it holds for all times. In the NEq case, the FDR in a strict sense is not available until the final equilibrium state is reached.  

It might be useful at this point to make this statement:   Although LRT has a more restricted specified condition than NEq, they both belong to open system set-ups, which is formally different from the close system set-up of eigenvalue thermalization. For a depiction of the differences between the two formulations, see, e.g., the Introduction of~\cite{HCSH}.\\

\noindent\underline{Weak coupling necessary for LRT, not required of NEq}

Weak coupling is assumed in LRT to ensure the weak disturbance and validity of the perturbative treatment from an otherwise free system dynamics. No detailed knowledge of the system and the bath is need. Thus, similar to conventional thermodynamics, the formalism can be applied to a wide class of configurations as long as the aforementioned assumptions are met. For NEq dynamics,  if  the system and its coupling with the bath are linear in nature,  Gaussian  dynamics is exactly solvable for arbitrarily strong coupling strength as long as it stays within the realm of dynamical stability. 
	
\subsection{FDRs in system-environment interaction with finite coupling} 

\noindent\underline{Final equilibrated state not a Gibbs form}  

In contrast to LRT where the system remains in a thermal state at the bath temperature,  in NEq dynamics,  the system undergoes  nonequilibrium evolution, and in most cases evolves to a final equilibrium state. In general, this equilibrium state is not a Gibbs (thermal) state unless the coupling strength between the system and the bath is vanishingly small.  In the open-system conceptual framework, at finite coupling strength,  there is a marked  difference between equilibration and thermalization.\\
	
\noindent\underline{End state temperature in NEq not the same as the bath.}  

Since in the NEq formalism the final state does not necessarily assume a Gibbs form, the system temperature, identified from the equilibrated reduced density matrix of the system, is at best `effective', because it  depends on the details of the system and the bath parameters, far from being universal. Only in the vanishing coupling limit will the reduced system's  temperature approach the initial temperature of the bath. However, the bath has evolved away from its initial thermal configuration, albeit the difference is not so significant if the system has fewer degrees of freedom.  Thus, only in this limit, where LRT operates, the system temperature, synonymous to the bath temperature, is  independent of the system and the bath.\\
	
\noindent\underline{Different proportionality factor}  

In LRT, the proportionality factor $\coth\beta\kappa/2$ in FDR depends on the system/bath temperature, while in the NEq formalism this factor depends on the initial temperature of the bath, which in general is not equal  to the effective temperature of the  final equilibrium state of the system, nor the temperature  of the bath, identified in its later evolution.
	
\subsection{FDR in a dynamical setting}
 
 In LRT, the FDR plays a rather passive, spectator role, relating the weak response of the system to an external agent's disturbance. By contrast,  in the NEQ formalism, FDR has a dynamical significance in  that it ensures that the energy flow   into the reduced system via the quantum fluctuations of the environment is balanced by the energy flow dispersed back to the environment in the form of  quantum radiation. This energy rate or power balance shows how the FDR regulates the environment's fluctuations  and the system's dissipation in a dynamical way. It also signifies the existence of an equilibrium state in this case. 
 
In summary, owing  to the requirement of self-consistency in the dynamics of the system and the bath, in the NEq theory the FDR at finite coupling strength cannot be given in an \textit{ad hoc} manner. Existence of the FDR is conditional upon the system reaching equilibrium,  which is not known a priori, but  determined by the dynamics of the reduced system. Beside, At finite coupling strength if it exists, this equilibrium state is not by default the Gibbs thermal state.

\section{Correlation-Propagation Relations (CPR) and  Non-Markovian mutual influence}\label{S:neire}

In the case that the system contains spatially-separated constituents, the FDR can be generalized to include the correlation-propagation relations (CPR) between the constituents. The quantum radiation coming from any constituent will propagate along the null cone reaching the other constituents, modifying their dynamics, and in turn affecting the ensuing radiations from each of them. This process will continue to multiply and reverberate, interlacing the constituents until the overall dynamics of the system finally settles down to equilibrium. This implies several important features: 1) any influence received by any one constituent depends on the state of all other constituents' motions at earlier moments.  This history-dependent influence is non-Markovian; 2) the retarded kernel of the environment describing the quantum radiation between constituents is connected to the correlation between the quantum fluctuations of the environment at the locations of the constituents; 3) The effect on the motion of each constituent due to the non-Markovian influence will then be nontrivially correlated with the action of the local quantum fluctuations of the environment. These features self-organize the motion of all constituents involved in accord at late times, and thus allow an emergent relation between the constituents. The CPR is absolutely crucial in keeping the balance of energy flows between the constituents of the system, and between the system and its environment. Without the consideration of this additional relation in tether with the FDR,  dynamical self-consistency cannot be sustained. Combination of these two sets of relations forms a generalized matrix FDR relation.  For further details on this point see~\cite{CPR2D,CPR4D}.

\section{Quantum radiation from an atom in 4D Minkowski spacetime}\label{S:etkjssd}

We now present a calculation to show the relation between vacuum fluctuations, in the form of quantum noise, and quantum radiation and the role played by the FDR. Consider a static Unruh-DeWitt detector, whose internal degree of freedom (idf) is modeled by a simple harmonic oscillator, coupled to a massless scalar field. We shall call this a harmonic atom, or simply the atom, to avoid confusion with a detector placed at a distance to measure emitted radiation. The field is initially in its thermal state, but the atom can assume any state. We assume the initial state of the total system is in a product form.  Since such a prepared initial state is usually not an energy eigenstate of the combined atom-field system, there will be energy fluctuations and exchanges between the atom and the field. The field fluctuations will drive the idf of the atom into random motion, which in turn emits quantum radiation  to the surrounding.  However, from the perspective of energy conservation, in particular if the field is initially in its vacuum state, then there is no energy reservoir that can sustain such  radiated energy. Thus, we expect that there is no quantum radiation from a stationary atom.  We shall reveal the energy budget of such a system and the role the FDRs play to balance the budget.

Earlier in~\cite{HCSH,CPR2D,CPR4D}, we have studied the mechanism of energy flow, or power balance, from the viewpoint of the atom. There, we find that the reactive force from  quantum radiation counters the motion of the atoms' idf,  driven by  quantum fluctuations of the field. As the idfs gradually settle down, the system reaching an equilibrium state, the energy flow funneled in from the field fluctuations is balanced by the dissipated energy flow from the idf of the atom due to the frictional reactive force. The FDRs of the field and the atom ensure this balance.   Here  focusing on the field, we explain how the energy balance can be reached even though no apparent conduit seem to exist to balance the radiated energy, thus highlighting the role of the FDRs in this `arrangement'.

\subsection{Quantum Langevin equation}\label{S:jhvsjdhs}

From the simultaneous set of Heisenberg equations of motion of the internal degree of freedom $\hat{Q}$ of the  harmonic atom moving along  a prescribed  trajectory $\mathbf{z}(t)$ and the massless scalar field $\hat{\phi}$ in $1+3$ unbounded Minkowski space,
\begin{align}
	\frac{d^{2}}{dt^{2}}\hat{Q}(t)+\omega_{0}^{2}\,\hat{Q}(t)&=\frac{e}{m}\,\hat{\phi}(\mathbf{z},t)\,,\label{E:ebkwe1}\\
	\Bigl(\frac{\partial^{2}}{\partial t^{2}}-\nabla^{2}\Bigr)\,\hat{\phi}(\mathbf{x},t)&=e\,\hat{Q}(t)\,\delta^{(3)}(\mathbf{x}-\mathbf{z})\,,\label{E:ebkwe2}
\end{align}
we can find the formal solution of the field operator as
\begin{equation}\label{E:bfkgskd}
	\hat{\phi}(\mathbf{x},t)=\hat{\phi}_{0,h}(\mathbf{x},t)+e\int\!d^{4}x'\;G_{0,R}^{(\phi)}(x,x')\,\hat{Q}(t')\,\delta^{(3)}(\mathbf{x}'-\mathbf{z}')\,,
\end{equation}
in which $e$ is the coupling constant, $\mathbf{z}'=\mathbf{z}(t')$, and $\hat{\phi}_{0,h}(\mathbf{x},t)$ is the homogeneous solution to the wave equation \eqref{E:ebkwe2}. The retarded Green's function $G_{0,R}^{(\phi)}(x^{\mu},x'^{\mu})$ of the free field, denoted by a subscript $0$ to distinguish it from the interacting case, satisfies
\begin{align}\label{E:rbdkhser}
	\Bigl(\frac{\partial^{2}}{\partial t^{2}}-\nabla^{2}\Bigr)\,G_{0,R}^{(\phi)}(x,x')&=\delta^{(4)}(x-x')\,,&&\text{with}&x^{\mu}&=(t,\mathbf{x})\,.
\end{align}
Eq.~\eqref{E:bfkgskd} clearly shows that the total field at the observation point $x$ comprises of the vacuum fluctuations of the free field and the radiation field produced by the quantum dynamics of the atom (further explanation later).

Putting \eqref{E:bfkgskd} back to \eqref{E:ebkwe1}, we arrive at
\begin{equation}\label{E:vdjfher}
	\frac{d^{2}}{dt^{2}}\hat{Q}(t)+\omega_{0}^{2}\,\hat{Q}(t)=\frac{e}{m}\,\hat{\phi}_{0,h}(\mathbf{z},t)+\frac{e^{2}}{m}\int\!dt'\;G_{0,R}^{(\phi)}(\mathbf{z},t;\mathbf{z}',t')\,\hat{Q}(t')\,.
\end{equation}
On the right hand side of this equation we see two source terms: The first term depicts vacuum field fluctuations  $\hat{\phi}_{0,h}(t,\mathbf{z})$  at the location $\mathbf{z}$ of the atom. It will induce stochastic motion in the internal degree of freedom of the atom. The second term contributing a nonlocal action in the equation of motion of the $\hat{Q}$ operator accounts for radiation from the atom. When the configuration involves only a single atom in the massless scalar field, this nonlocal term reduces to two local contributions: one accounts for frequency renormalization and the other is a local damping term, such that \eqref{E:vdjfher} takes a rather simple form
\begin{equation}\label{E:nreonsds}
	\frac{d^{2}}{dt^{2}}\hat{Q}(t)+2\gamma\,\dot{\hat{Q}}+\omega^{2}\,\hat{Q}(t)=\frac{e}{m}\,\hat{\phi}_{0,h}(\mathbf{z},t)\,,
\end{equation} 
where $\gamma=e^{2}/8\pi m$ is the damping constant, whose inverse gives the time scale of relaxation, and $\omega$ is the physical frequency, which includes the bare frequency and the renormalization correction. Thus, we see the quantum radiation from the atom will introduce damping in the equation of motion of the idf of the atom, this is the reactive force of quantum radiation we referred to earlier. Eq.~\eqref{E:nreonsds} thus says that the internal degree of freedom of the atom essentially behaves like a driven damped oscillator ~\cite{HCSH,HHPRD} --  driven by the quantum fluctuations of the field, but is damped by the reactive force of quantum radiation.

The solution to \eqref{E:vdjfher} or \eqref{E:nreonsds} is given by
\begin{equation}\label{E:rutherwo1}
	\hat{Q}(t)=\hat{Q}_{h}(t)+\frac{e}{m}\int_{0}^{t}\!ds\;G_{R}^{(Q)}(t-s)\,\hat{\phi}_{0,h}(\mathbf{z}_{s},s)\,.
\end{equation}
with $\mathbf{z}_{s}=\mathbf{z}(s)$. Here, the homogeneous part $\hat{Q}_{h}(t)$ satisfies 
\begin{equation}
	\frac{d^{2}}{dt^{2}}\hat{Q}_{h}(t)+\omega_{0}^{2}\,\hat{Q}_{h}(t)-\frac{e^{2}}{m}\int^{t}_{0}\!dt'\;G_{0,R}^{(\phi)}(\mathbf{z},t;\mathbf{z}',t')\,\hat{Q}_{h}(t')=\ddot{\hat{Q}}_{h}(t)+2\gamma\,\dot{\hat{Q}}_{h}+\omega^{2}\,\hat{Q}_{h}(t)=0\,.
\end{equation}
The corresponding Green's function $G_{R}^{(Q)}(t-t')$ for the internal degree of freedom of the atom has a Fourier transform of the form
\begin{align}
	\overline{G}_{R}^{(Q)}(\kappa)&=\frac{1}{-\kappa^{2}+\omega-i\,2\gamma\,\kappa}=\frac{1}{-\kappa^{2}+\omega_{0}^{2}-\dfrac{e^{2}}{m}\,\overline{G}_{0,R}^{(\phi)}(\mathbf{0};\kappa)}\,.
\end{align}
The Fourier transformation of a function $f(t)$ is defined by
\begin{equation}
	\overline{f}(\kappa)=\int_{-\infty}^{\infty}\!dt\;f(t)\,e^{+i\kappa t}\,.
\end{equation}
{From \eqref{E:rutherwo1}, we see the dynamics of the internal degree of freedom $\hat{Q}$ of the atom contains two stochastic 
 components: One comes from the homogeneous part of the solution, denoted by $\hat{Q}_{h}$, which depends on the initial conditions. This part is intrinsic in the sense that it is still present when the coupling is turned off, but will decay with time due to damping when the interaction is turned on. The other component is the inhomogeneous part in \eqref{E:rutherwo1}, induced by the quantum fluctuations of the free field. The joint random motion of the idf, both of quantum origin, will give off quantum radiation to the atom's surrounding, as can be seen from the second term on the right hand side of \eqref{E:bfkgskd}.}

If we want to investigate the properties of the field, Eq.~\eqref{E:bfkgskd} is an easier start. Since it is driven by the full $\hat{Q}$ operator, we may substitute \eqref{E:rutherwo1} into \eqref{E:bfkgskd} and write the interacting field \eqref{E:bfkgskd} as
\begin{align}\label{E:zauubsds3}
	\hat{\phi}(\mathbf{x},t)&=\hat{\phi}_{0,h}(\mathbf{x},t)+\frac{e^{2}}{m}\int\!dt'\;G_{0,R}^{(\phi)}(\mathbf{x},t;\mathbf{z}',t')\int_{0}^{t'}\!ds\;G_{R}^{(Q)}(t'-s)\,\hat{\phi}_{0,h}(\mathbf{z}_{s},s)\notag\\
	&\qquad\qquad\qquad\qquad\qquad\qquad\qquad\qquad\qquad\qquad+e\int_{0}^{t}\!dt'\;G_{0,R}^{(\phi)}(\mathbf{x},t;\mathbf{z}',t')\,\hat{Q}_{h}(t')\\
	&=\hat{\phi}_{0,h}(\mathbf{x},t)+\hat{\phi}_{\textsc{r}}(\mathbf{x},t)+\hat{\phi}_{\textsc{tr}}(\mathbf{x},t)\,,
\end{align}
where $\hat{\phi}_{\textsc{r}}$ in the second term on the right hand side is the radiation field caused by the driven stochastic motion of the idf of the atom, and the third term on the right hand side, denoted by $\hat{\phi}_{\textsc{tr}}$ is the transient radiation field from the homogeneous solution of $\hat{Q}$. The Hadamard function of the interacting field is readily given by
\begin{align}
	G_{H}^{(\phi)}(x,x')&=\frac{1}{2}\,\langle\bigl\{\hat{\phi}(x),\hat{\phi}(x')\bigr\}\rangle\notag\\
	&=\frac{1}{2}\,\langle\bigl\{\hat{\phi}_{0,h}(x),\hat{\phi}_{0,h}(x')\bigr\}\rangle+\frac{1}{2}\,\langle\bigl\{\hat{\phi}_{0,h}(x),\hat{\phi}_{\textsc{r}}(x')\bigr\}\rangle+\frac{1}{2}\,\langle\bigl\{\hat{\phi}_{\textsc{r}}(x),\hat{\phi}_{0,h}(x')\bigr\}\rangle\notag\\
	&\qquad\qquad\qquad\qquad\qquad\qquad+\frac{1}{2}\,\langle\bigl\{\hat{\phi}_{\textsc{r}}(x),\hat{\phi}_{\textsc{r}}(x')\bigr\}\rangle+\frac{1}{2}\,\langle\bigl\{\hat{\phi}_{\textsc{tr}}(x),\hat{\phi}_{\textsc{tr}}(x')\bigr\}\rangle\,.\label{E:dbfksdj}
\end{align}
We see that on the right hand side the first term of \eqref{E:dbfksdj} results from the free field at the observation point, and the fourth term {purely from the radiation field due to the induced stochastic motion of the internal degree of freedom}. The second and third terms are of interest because they represent the interference between the free field and the radiation field at the observation point. The last term in \eqref{E:khgrsdfs} does not contribute at late times because the homogeneous solution of the driven and damped oscillator exponentially decays with time, and for a given observation point, the radiation due to this component does not interfere with the local vacuum field and decreases with time.

{Putting in the explicit expressions of $\hat{\phi}(x)$ into \eqref{E:dbfksdj}, we obtain}
\begin{align}
	G_{H}^{(\phi)}(x,x')&=G_{0,H}^{(\phi)}(x,x')+\frac{e^{2}}{m}\int_{0}^{t'}\!ds'_{1}\;G_{0,R}^{(\phi)}(\mathbf{x}',t';\mathbf{z}'_{1},s'_{1})\int_{0}^{s'_{1}}\!ds'_{2}\;G_{R}^{(Q)}(s'_{1}-s'_{2})\,G_{0,H}^{(\phi)}(\mathbf{x},t;\mathbf{z}'_{2},s'_{2})\notag\\
	&\qquad\qquad\qquad\;+\frac{e^{2}}{m}\int_{0}^{t}\!ds_{1}\;G_{0,R}^{(\phi)}(\mathbf{x},t;\mathbf{z}_{1},s_{1})\int_{0}^{s_{1}}\!ds_{2}\;G_{R}^{(Q)}(s_{1}-s_{2})\,G_{0,H}^{(\phi)}(\mathbf{x}',t';\mathbf{z}_{2},s_{2})\notag\\
	&\qquad+\frac{e^{4}}{m^{2}}\int_{0}^{t}\!ds_{1}\int_{0}^{t'}\!ds'_{1}\;G_{0,R}^{(\phi)}(\mathbf{x},t;\mathbf{z}_{1},s_{1})G_{0,R}^{(\phi)}(\mathbf{x}',t';\mathbf{z}'_{1},s'_{1})\notag\\
	&\qquad\qquad\qquad\times\int_{0}^{s_{1}}\!ds_{2}\int_{0}^{s'_{1}}\!ds'_{2}\;G_{R}^{(Q)}(s_{1}-s_{2})G_{R}^{(Q)}(s'_{1}-s'_{2})\,G_{0,H}^{(\phi)}(\mathbf{z}^{\vphantom{!}}_{2},s^{\vphantom{!}}_{2};\notag\mathbf{z}'_{2},s'_{2})\notag\\
	&\qquad+\frac{e^{2}}{2}\int_{0}^{t}\!ds\int_{0}^{t'}\!ds'\;G_{0,R}^{(\phi)}(\mathbf{x},t;\mathbf{z}_{s},s)G_{0,R}^{(\phi)}(\mathbf{x},t;\mathbf{z}_{s}',s')\,\langle\bigl\{\hat{Q}_{h}(s),\hat{Q}_{h}(s')\bigr\}\rangle\,,\label{E:khgrsdfs}
\end{align}
where $G_{0,H}^{(\phi)}(x,x')$ is the Hadamard function of the free field
\begin{equation}
	G_{0,H}^{(\phi)}(x,x')=\frac{1}{2}\,\langle\bigl\{\hat{\phi}_{0,h}(x),\hat{\phi}_{0,h}(x')\bigr\}\rangle\,,
\end{equation}
and we have assumed that the initial state of the atom's internal degree of freedom does not have cross correlation in the canonical variables.

The correlation function \eqref{E:khgrsdfs} will be used to construct the stress-energy tensor of the interacting field $\hat{\phi}$, based on which we will compute the late-time energy flow of the scalar field at an observation point far away from the atom.

\subsection{A stationary atom}

We consider the simplest case of an atom placed at a fixed location so that $\mathbf{z}$ is independent of time. Let $\mathbf{r}_{1}=\mathbf{x}'-\mathbf{z}$, $\mathbf{r}_{2}=\mathbf{x}-\mathbf{z}$. For $t$, $t'\gg\gamma^{-1}$ and $t$, $t'\gg r_{1}$, $r_{2}$, we can use the fluctuation-dissipation relations of the atom and the free field
\begin{align}\label{E:fbgrusyer}
	\overline{G}_{H}^{(Q)}&=\coth\frac{\beta\kappa}{2}\,\operatorname{Im}\overline{G}_{R}^{(Q)}(\kappa)\,,&\overline{G}_{0,H}^{(\phi)}&=\coth\frac{\beta\kappa}{2}\,\operatorname{Im}\overline{G}_{0,R}^{(\phi)}(\kappa)\,,
\end{align}
in the expression
\begin{equation}\label{E:dbferysd}
	\overline{G}_{0,H}^{(\phi)}(\mathbf{0};\kappa)\,\overline{G}_{R}^{(Q)}(\kappa)\,\overline{G}_{R}^{(Q)*}(\kappa)=\frac{m}{e^{2}}\frac{\overline{G}_{0,H}^{(\phi)}(\mathbf{0};\kappa)}{\operatorname{Im}\overline{G}_{0,R}^{(\phi)}(\mathbf{0};\kappa)}\,\operatorname{Im}\overline{G}_{R}^{(Q)}(\kappa)=\frac{m}{e^{2}}\,\coth\frac{\beta\kappa}{2}\,\operatorname{Im}\overline{G}_{R}^{(Q)}(\kappa)\,,
\end{equation} 
to greatly simplify the Hadamard function \eqref{E:khgrsdfs} of the interacting field,
\begin{align}\label{E:yreygte}
	G_{H}^{(\phi)}(x,x')&=G_{0,H}^{(\phi)}(x,x')+\frac{e^{2}}{m}\int_{-\infty}^{\infty}\!\frac{d\kappa}{2\pi}\;\overline{G}_{0,H}^{(\phi)}(\mathbf{r}_{2};\kappa)\,\overline{G}_{0,R}^{(\phi)*}(\mathbf{r}_{1};\kappa)\,\overline{G}_{R}^{(Q)*}(\kappa)\,e^{-i\kappa(t-t')}\notag\\
	&\qquad\qquad\qquad\,+\frac{e^{2}}{m}\int_{-\infty}^{\infty}\!\frac{d\kappa}{2\pi}\;\overline{G}_{0,H}^{(\phi)}(\mathbf{r}_{1};\kappa)\,\overline{G}_{0,R}^{(\phi)}(\mathbf{r}_{2};\kappa)\,\overline{G}_{R}^{(Q)}(\kappa)\,e^{-i\kappa(t-t')}\notag\\
	&\quad+\frac{e^{2}}{m}\int_{-\infty}^{\infty}\!\frac{d\kappa}{2\pi}\;\coth\frac{\beta\kappa}{2}\,\operatorname{Im}\overline{G}_{R}^{(Q)}(\kappa)\,\overline{G}_{0,R}^{(\phi)}(\mathbf{r}_{2};\kappa)\,\overline{G}_{0,R}^{(\phi)*}(\mathbf{r}_{1};\kappa)\,e^{-i\kappa(t-t')}\,,
\end{align}
The last term in \eqref{E:khgrsdfs} has been discarded. Application of the FDR in \eqref{E:dbferysd} reduces the contribution purely from the radiation to a form that resembles those from the interference terms. This may not be a surprise on account that the late-time dynamics of the internal degree of freedom of the atom is predominantly governed by the quantum fluctuations of the free field in close proximity of the atom. A quick check shows that Eq.~\eqref{E:yreygte} satisfies the invariant property of the Hadamard function under the exchange of the coordinate points $x^{\mu}=(t,\mathbf{x})$ and $x'^{\mu}=(t',\mathbf{x}')$. Also note that in \eqref{E:fbgrusyer}, we have assumed that the field is initially in a thermal state of temperature $\beta^{-1}$, so the FDR of the free field is formulated for its initial state. By contrast, the FDR of the atom is obtained when the idf of the atom has relaxed to equilibrium after interacting with the field, despite their formal identity. In the case that the field is initially in a vacuum state, i.e., $\beta\to\infty$, just replace the factor $\coth\beta\kappa/2$ by the sign function $\operatorname{sgn}(\kappa)$.

\subsection{Stress-energy tensor of field}

The classical stress-energy tensor of a massless scalar field is given by
\begin{equation}
	T_{\mu\nu}(x)=\phi_{,\mu}(x)\phi_{,\nu}(x)-\frac{1}{2}\,g_{\mu\nu}g^{\alpha\beta}\phi_{,\alpha}(x)\phi_{,\beta}(x)\,.
\end{equation}
We will use the correlation function \eqref{E:yreygte} to find the expectation value of the normal-ordered  stress-energy tensor operator $\hat{T}_{\mu\nu}(x)$ by
\begin{equation}
	\langle:\hat{T}_{\mu\nu}(x):\rangle=\lim_{x'\to x}\biggl\{\frac{\partial^{2}}{\partial x^{\mu}\partial x'^{\nu}}-\frac{1}{2}\,g_{\mu\nu}g^{\alpha\beta}\,\frac{\partial^{2}}{\partial x^{\alpha}\partial x'^{\beta}}\biggr\}\Bigl[G_{H}^{(\phi)}(x,x')-G_{0,H}^{(\phi)}(x,x')\Bigr]\,.
\end{equation}
The radiated power far away from the atom is then given by
\begin{equation}\label{E:rbguder}
	\frac{dW_{\textsc{rad}}}{d\tau}=-\int\!d\Omega\;r^{2}n^{\mu}\langle:\hat{T}_{\mu\nu}(x):\rangle\,v^{\nu}(\tau_{-})\,,
\end{equation}
where $v^{\mu}(\tau_{-})$ is the four-velocity of the (motion of the center of mass of the) atom at the retarded time $\tau_{-}$ from the observation point $x$, and $n^{\mu}$ is a spacelike unit radial vector, so that the spatial distance $r$ between the observation point and the atom at the retarded time is given by
\begin{equation}
	r=n^{\mu}\bigl[x_{\mu}-z_{\mu}(\tau_{-})\bigr]\,,
\end{equation}
that is, the radius of the spherical shell passing through the field point $x^{\mu}$ with the origin at the retarded source point $z_{\mu}(\tau_{-})$ of the atom. In the case of a static atom, the location $z^{\mu}$ is fixed, so the four-velocity is a constant timelike unit vector along the time axis, whereby \eqref{E:rbguder} becomes
\begin{equation}
	\frac{dW_{\textsc{rad}}}{d\tau}=-\int\!d\Omega\;r^{2}\,\langle : T_{rt}(x):\rangle=-\lim_{x'\to x}\int\!d\Omega\;r^{2}\,\frac{\partial^{2}}{\partial r\partial t'}\Bigl[G_{H}^{(\phi)}(x,x')-G_{0,H}^{(\phi)}(x,x')\Bigr]\,,
\end{equation}
where $d\Omega$ is the solid angle subtended over the spherical shell.

The coincidence limit of the  derivatives
\begin{align}
	\lim_{x'\to x}\frac{\partial^{2}}{\partial r\partial t'}\Bigl[G_{H}^{(\phi)}(x,x')-G_{0,H}^{(\phi)}(x,x')\Bigr]
\end{align}
at late times is given by
\begin{align}
	&\quad\int_{-\infty}^{\infty}\!\frac{d\kappa}{2\pi}\;\biggl\{i\kappa\,\Bigl[\kappa\,\coth\frac{\beta\kappa}{2}\operatorname{Re}\overline{G}_{0,R}^{(\phi)}(\mathbf{r};\kappa)-\frac{1}{r}\,\overline{G}_{0,H}^{(\phi)}(\mathbf{r};\kappa)\Bigr]\,\overline{G}_{0,R}^{(\phi)*}(\mathbf{r};\kappa)\,\overline{G}_{R}^{(Q)*}(\kappa)\biggr.\notag\\
	&\qquad\qquad\qquad\qquad+i\kappa\,\Bigl(i\kappa-\frac{1}{r}\Bigr)\,\overline{G}_{0,H}^{(\phi)}(\mathbf{r};\kappa)\,\overline{G}_{0,R}^{(\phi)}(\mathbf{r};\kappa)\,\overline{G}_{R}^{(Q)}(\kappa)\notag\\
	&\qquad\qquad\qquad\qquad\qquad\quad+\biggl.i\,\kappa\,\Bigl(i\kappa-\frac{1}{r}\Bigr)\,\coth\frac{\beta\kappa}{2}\,\operatorname{Im}\overline{G}_{R}^{(Q)}(\kappa)\,\overline{G}_{0,R}^{(\phi)}(\mathbf{r};\kappa)\,\overline{G}_{0,R}^{(\phi)*}(\mathbf{r};\kappa)\biggr\}\,.\label{E:dkbehrs}
\end{align}
Since we are interested in the flux at large distance from the atom, we keep only the far-field component in \eqref{E:dkbehrs}, and obtain
\begin{align}
	&\quad\int_{-\infty}^{\infty}\!\frac{d\kappa}{2\pi}\;\biggl\{i\kappa^{2}\,\coth\frac{\beta\kappa}{2}\,\operatorname{Re}\overline{G}_{0,R}^{(\phi)}(\mathbf{r};\kappa)\,\overline{G}_{0,R}^{(\phi)*}(\mathbf{r};\kappa)\,\overline{G}_{R}^{(Q)*}(\kappa)-\kappa^{2}\,\overline{G}_{0,H}^{(\phi)}(\mathbf{r};\kappa)\,\overline{G}_{0,R}^{(\phi)}(\mathbf{r};\kappa)\,\overline{G}_{R}^{(Q)}(\kappa)\biggr.\notag\\
	&\qquad\qquad\qquad\qquad\qquad\qquad-\biggl.\kappa^{2}\,\coth\frac{\beta\kappa}{2}\,\operatorname{Im}\overline{G}_{R}^{(Q)}(\kappa)\,\overline{G}_{0,R}^{(\phi)}(\mathbf{r};\kappa)\,\overline{G}_{0,R}^{(\phi)*}(\mathbf{r};\kappa)\biggr\}\notag\\
	&=\int_{-\infty}^{\infty}\!\frac{d\kappa}{2\pi}\;\kappa^{2}\,\coth\frac{\beta\kappa}{2}\,\biggl\{i\,\operatorname{Re}\overline{G}_{0,R}^{(\phi)}(\mathbf{r};\kappa)\,\overline{G}_{0,R}^{(\phi)*}(\mathbf{r};\kappa)\,\overline{G}_{R}^{(Q)*}(\kappa)-\operatorname{Im}\overline{G}_{0,R}^{(\phi)}(\mathbf{r};\kappa)\,\overline{G}_{0,R}^{(\phi)}(\mathbf{r};\kappa)\,\overline{G}_{R}^{(Q)}(\kappa)\biggr.\notag\\
	&\qquad\qquad\qquad\qquad\qquad\qquad+\biggl.i\,\overline{G}_{R}^{(Q)}(\kappa)\,\overline{G}_{0,R}^{(\phi)}(\mathbf{r};\kappa)\,\overline{G}_{0,R}^{(\phi)*}(\mathbf{r};\kappa)\biggr\}\,.\label{E:pawmrsrj}\\
	\intertext{Here we have used the FDR of the free field to re-write \eqref{E:pawmrsrj}. It then can be further reduced to}
	&=\int_{-\infty}^{\infty}\!\frac{d\kappa}{2\pi}\;\kappa^{2}\,\coth\frac{\beta\kappa}{2}\biggl\{-i\,\operatorname{Re}\overline{G}_{0,R}^{(\phi)}(\mathbf{r};\kappa)\,\overline{G}_{0,R}^{(\phi)}(\mathbf{r};\kappa)\,\overline{G}_{R}^{(Q)}(\kappa)-\operatorname{Im}\overline{G}_{0,R}^{(\phi)}(\mathbf{r};\kappa)\,\overline{G}_{0,R}^{(\phi)}(\mathbf{r};\kappa)\,\overline{G}_{R}^{(Q)}(\kappa)\biggr.\notag\\
	&\qquad\qquad\qquad\qquad\qquad\qquad+\biggl.i\,\overline{G}_{R}^{(Q)}(\kappa)\,\overline{G}_{0,R}^{(\phi)}(\mathbf{r};\kappa)\,\overline{G}_{0,R}^{(\phi)*}(\mathbf{r};\kappa)\biggr\}\notag\\
	&=\int_{-\infty}^{\infty}\!\frac{d\kappa}{2\pi}\;\kappa^{2}\,\coth\frac{\beta\kappa}{2}\biggl\{-i\,\overline{G}_{0,R}^{(\phi)*}(\mathbf{r};\kappa)\,\overline{G}_{0,R}^{(\phi)}(\mathbf{r};\kappa)\,\overline{G}_{R}^{(Q)}(\kappa)+i\,\overline{G}_{R}^{(Q)}(\kappa)\,\overline{G}_{0,R}^{(\phi)}(\mathbf{r};\kappa)\,\overline{G}_{0,R}^{(\phi)*}(\mathbf{r};\kappa)\biggr\}\,,\label{E:bgrhdsdf}
\end{align}
where we have used the property that $\operatorname{Im}\overline{G}_{R}^{(Q)}(\kappa)$ is odd in $\kappa$, $\operatorname{Re}\overline{G}_{R}^{(Q)}(\kappa)$ even in $\kappa$ and $\overline{G}_{R}(-\kappa)=\overline{G}_{R}^{*}(\kappa)$. Eq.~\eqref{E:bgrhdsdf} gives a vanishing result. Thus we find
\begin{equation}\label{E:dfkdhd}
	\frac{dW_{\textsc{rad}}}{d\tau}=0\,,
\end{equation}
that is, there is no net radiated energy at large distance from the atom at late times $t\gg r$. 

\subsection{Energy flow balance}

This result is as expected.  But this immediately prompts a question about how the radiated energy due to the stochastic motion of the internal degrees of freedom of the atom gets canceled? To track down the inner workings it will be more instructive to write the net radiated power \eqref{E:dfkdhd} as a sum of $P_{r}$ and $P_{\times}$, which are the energy flow associated with the second and the first terms inside the curly brackets, respectively,  on the right hand side of \eqref{E:bgrhdsdf}. They can further be reduced to
\begin{align}
	P_{r}&=-i\,\frac{e^{2}}{m}\int_{-\infty}^{\infty}\!\frac{d\kappa}{2\pi}\;\frac{\kappa^{2}}{4\pi}\,\coth\frac{\beta\kappa}{2}\,\overline{G}_{R}^{(Q)}(\kappa)=+\frac{e^{2}}{m}\int_{-\infty}^{\infty}\!\frac{d\kappa}{2\pi}\;\frac{\kappa^{2}}{4\pi}\,\coth\frac{\beta\kappa}{2}\,\operatorname{Im}\overline{G}_{R}^{(Q)}(\kappa)\,,\label{E:bgks1}\\
	P_{\times}&=-\frac{e^{2}}{m}\int_{-\infty}^{\infty}\!\frac{d\kappa}{2\pi}\;\frac{\kappa^{2}}{4\pi}\,\coth\frac{\beta\kappa}{2}\,\operatorname{Im}\overline{G}_{R}^{(Q)}(\kappa)\,.\label{E:bgks2}
\end{align}
The radiation generated by the {induced stochastic} motion of the atom's internal degree of freedom gives a power $P_{r}$ at large distance from the atom at late times. This comes solely  from the far-field component of the radiated field, and it gives an outward energy flow. However, we note that the radiated field also interfere with the local vacuum fluctuations. This results at infinity in a net incoming flux of the same magnitude  which cancels the aforementioned outgoing flux -- an inertial observer in the distance will not measure any net energy flow. The generalized FDR, in particular the CPR, of the free field used in \eqref{E:pawmrsrj} tells us that at an observation point far away from the atom, the radiation field is correlated with the vacuum field at that point in a nontrivial way, according to Sec.~\ref{S:neire} that their interference can concoct up an incoming flux to  enable a perfect  balance in the energy flow. Physically this is a consequence of energy conservation. In particular,  if the surrounding field is initially in the vacuum state, and if there is no external agent to supply energy, then there is no source/drain of energy to support a net energy flow in or out when the whole system settles down into equilibrium.

It will be more interesting to compare $P_{r}$ and $P_{\times}$ with the power  $P_{\gamma}$ dissipated by the atom and the power $P_{\xi}$ supplied to the atom by the local vacuum fluctuations of the free field around the atom, at late times,  
\begin{equation}
	P_{\gamma}=-\frac{e^{2}}{m}\int^{\infty}_{-\infty}\!\frac{d\kappa}{2\pi}\;\kappa\,\operatorname{Im}\overline{G}^{(Q)}_{R}(\kappa)\,\overline{G}_{H}^{(\phi)}(\mathbf{0};\kappa)=-\frac{e^{2}}{m}\int_{-\infty}^{\infty}\!\frac{d\kappa}{2\pi}\;\frac{\kappa^{2}}{4\pi}\,\coth\frac{\beta\kappa}{2}\,\operatorname{Im}\overline{G}_{R}^{(Q)}(\kappa)=-P_{\xi}\,,\label{E:bgks3}
\end{equation}
where we have used the FDR of the atom \eqref{E:dbferysd}, and the fact that
\begin{equation}
	\overline{G}_{H}^{(\phi)}(\mathbf{0};\kappa)=\lim_{r\to0}\overline{G}_{H}^{(\phi)}(\mathbf{r};\kappa)=\lim_{r\to0}\coth\frac{\beta\kappa}{2}\,\frac{\sin\kappa r}{4\pi r}=\frac{\kappa}{4\pi}\,\coth\frac{\beta\kappa}{2}\,.
\end{equation}
Eq.~\eqref{E:bgks3} explicitly shows that when the motion of the internal degrees of freedom of the atom reach  equilibrium, the energy flow out of the atom due to the dissipative reactive force of quantum radiation  is balanced by the power input from the local vacuum fluctuations of the free field.   Therefore,  comparison of Eqs.~\eqref{E:bgks1}--\eqref{E:bgks3} tells us that this dissipated energy is radiated outward to infinity, i.e.  $P_{\gamma}=P_{r}$. If we inadvertently ignore the contribution from the interference between the radiated field and the vacuum at the faraway location, then we will have a net energy output to infinity, which violates energy conservation. The interference contribution results in an inward energy flow which on one hand cancels the outgoing radiated energy and on the other hand replenishes the energy that is dumped into the atom from the vacuum field around it, that is, $P_{\times}=P_{\xi}$. Thus the energy flow, from the viewpoint of either the atom or the field, is perfectly balanced at late times. Performing these cross-checks enables us to present an integrated and comprehensive picture of how the energy transfers from the atom to the surrounding field and back to the atom,  meeting the stringent self-consistency constraint conditions ingrained in the FDRs-CPRs.\\

\acknowledgments{JTH acknowledges the hospitality at the MCFP and JQI of the University of Maryland where this work is carried out.}



\end{document}